\documentclass[12pt]{iopart}
\usepackage{graphicx}
\usepackage{mathtext}

\begin{document}

\newcommand {\la} {\left\langle}
\newcommand {\ra} {\right\rangle}

\title[Scenarios of washing-out of localized patterns under frozen parametric disorder]{Two scenarios of advective washing-out of localized convective patterns under frozen parametric disorder}

\author{Denis S Goldobin}
\address{Institute of Continuous Media Mechanics UB RAS,
         Perm 614013, Russia}
\address{Department of Theoretical Physics, Perm State University,
	     Perm 614990, Russia}
\ead{Denis.Goldobin@gmail.com}

\begin{abstract}
The effect of spatial localization of states in distributed parameter systems under frozen parametric disorder is well known as the Anderson localization and thoroughly studied for the Schr\"odinger equation and linear dissipation-free wave equations. Some similar ({\it or} mimicking) phenomena can occur in dissipative systems such as the thermal convection ones. Specifically, many of these dissipative systems are governed by a modified Kuramoto--Sivashinsky equation, where the frozen spatial disorder of parameters has been reported to lead to excitation of localized patterns. Imposed advection in the modified Kuramoto--Sivashinsky equation can affect the localized patterns in a nontrivial way; it changes the localization properties and suppresses the pattern. The latter effect is considered in this paper by means of both numerical simulation and model reduction, which turns out to be useful for a comprehensive understanding of the bifurcation scenarios in the system. Two possible bifurcation scenarios of advective suppression (``washing-out'') of localized patterns are revealed and characterised.
\end{abstract}


\vspace{2pc}
\noindent{\it Special Issue}: Article preparation, IOP journals\\
\submitto{\it \PS}
\maketitle


\section{Introduction}
The phenomenon of localization of linear waves in media with frozen disorder of distributed parameters---An\-der\-son localization~\cite{Anderson-1958,Abrahams-etal-1979}---is one of the most distinguishable findings in mathematical physics in the 20th century. The phenomenon has drawn attention and earned its name after the discovery of the effect in quantum mechanics~\cite{Anderson-1958}. Later on, the Anderson localization has been widely studied in quantum optics ({\it e.g.}, see~\cite{John-1987,Schwartz-etal-2007}) and diverse branches of classical physics; for instance, in wave optics~\cite{Klyatskin-2005} and acoustics~\cite{Maynard-2001}. The phenomenon has been well understood mathematically for the Schr\"odinger equation and linear dissipation-free wave equations~\cite{Abrahams-etal-1979,Froehlich-Spencer-1984,Lifshitz-Gredeskul-Pastur-1988,Gredeskul-Kivshar-1992}.

The absence of dissipation in the system is essential for the very concept of the Anderson localization. However, for an active/dissipative medium with frozen parametric disorder the mathematical formulation of some problems can be formally very similar to that of the Anderson localization~\cite{Goldobin-Shklyaeva-2013}. Nonetheless, to the author's knowledge, the localization-related aspects in the behaviour of dissipative systems with frozen parametric disorder have been studied only in~\cite{Goldobin-Shklyaeva-2013,Goldobin-Shklyaeva-2009,Goldobin-2010}, while few other authors considered the effect of disorder on the linear instability threshold of the base state of the system~\cite{Hammele-Schuler-Zimmermann-2006,Limia-Kofane-2016}.

\subsection*{Localized patterns in dissipative systems}
\begin{figure}[!b]
\center{
\includegraphics[width=0.40\textwidth]%
 {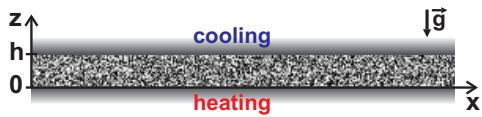}
}
\caption{Fluid-saturated porous layer heated from below and the coordinate frame}
\label{fig1}
\end{figure}

A fluid-saturated horizontal porous layer with nearly insulating thermal boundaries heated from below (see Fig.~\ref{fig1}) is an example of the system where frozen disorder of distributed parameters is physically plausible; it can be related to the imperfectness of the homogeneity of the solid matrix packing and applied heat flux. For the case of a thin layer, the temperature perturbation is nearly uniform along the vertical coordinate $z$ (Fig.~\ref{fig1}), $\theta=\theta(x,y,t)$, and its dynamics is governed by equation (see~\cite{Goldobin-Shklyaeva-BR-2008})
\[
\dot\theta+\vec{u}\cdot\nabla\theta+\Delta^2\theta
 -\nabla\cdot\big(\nabla\theta(\nabla\theta)^2-q(x,y)\nabla\theta\big)=0\,,
\]
where $q(x,y)$ is the relative deviation of the local Rayleigh--Darcy number from the critical value of the homogeneous layer and $\vec{u}(x,y)$ is an imposed advective through-flow uniquely determined by the lateral boundary conditions and the conservation law $\nabla\cdot\vec{u}=0$. Positive $q(x,y)$ means the local excess over the instability threshold of the uniform system; $q$ can be referred to as ``supercriticality''.

\begin{figure*}[!t]
\center{
\includegraphics[width=0.60\textwidth]%
{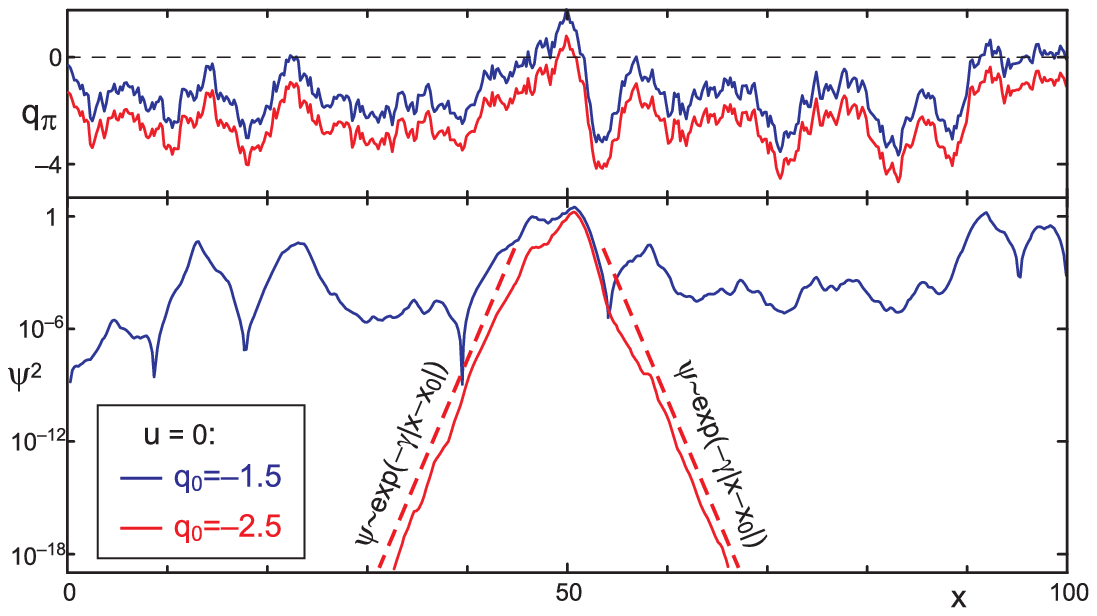}\\[5pt]
\includegraphics[width=0.60\textwidth]%
{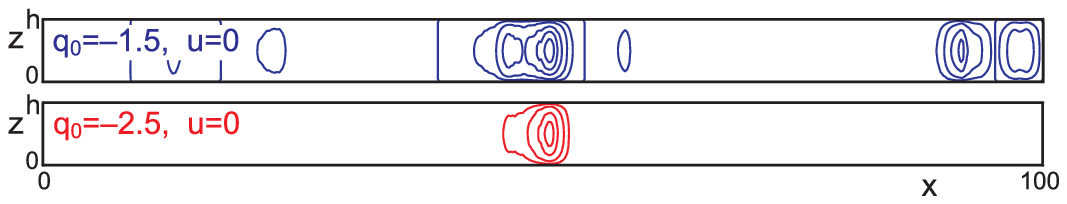}
}
\caption{Localized pattern and convective flows for $u=0$}
\label{fig2}
\end{figure*}

We will consider the case of a system uniform along the $y$-coordinate. Then the governing equation turns into a modified Kuramoto--Sivashinsky equation
\begin{equation}
\dot{\theta}=\left(-u\theta-\theta_{xxx}-q(x)\,\theta_x+(\theta_x)^3\right)_x\,,
\label{eq2-01}
\end{equation}
where $u$ is the $x$-component of the imposed advection (through-flow) velocity and must be constant. Here, we would like to emphasize that Eq.~(\ref{eq2-01}) is a typical governing equation for the pattern formation in systems with symmetry $\theta\leftrightarrow-\theta$ (this symmetry emerges also in the systems where $\theta$ is an amplitude of a mode oscillating in space or time); for instance, see~\cite{Knobloch-1990,Shtilman-Sivashinsky-1991,Hoyle-1998,Schoepf-Zimmermann-1989,Schoepf-Zimmermann-1993}.

The frozen parametric disorder is represented by a $\delta$-correlated Gaussian noise:
\begin{equation}
\begin{array}{c}
q(x)=q_0+\xi(x)\,,
\\[5pt]
\la\xi(x)\ra=0\,,\quad
\la\xi(x)\,\xi(x')\ra=2\sigma^2\delta(x-x')\,,
\end{array}
\label{eq2-02}
\end{equation}
where $q_0$ is the mean supercriticality, $\sigma$ is the noise strength. The noise strength $\sigma$ may be set to $1$ by means of the rescaling $(x,t,q)\to(\sigma^{-2/3}x,\sigma^{-8/3}t,\sigma^{4/3}q)$.

In~\cite{Goldobin-Shklyaeva-BR-2008}, the fluid velocity field of large-scale convective currents in a porous layer has been reported;
\begin{equation}
\vec{v}=\frac{\partial\Psi}{\partial z}\vec{e}_x
-\frac{\partial\Psi}{\partial x}\vec{e}_z\,,
\qquad
\Psi=f(z)\,\psi(x,t)\,,
\label{eq2-03}
\end{equation}
where $\psi(x,t)=\theta_x(x,t)$ is the stream function amplitude, $f(z)=3\sqrt{35}\,\sigma^{-1}h^{-3}z(h-z)$, $h$ is the layer thickness. Notice, for different convective systems function $f(z)$ will be different, but the relation $\psi(x,t)=\theta_x(x,t)$ will be preserved.

As demonstrated in~\cite{Goldobin-Shklyaeva-2013}, for $u=0$ and $q_0<0$ only time-independent patterns are stable. For moderate negative values of $q_0$ the system state turns into a set of separate localized convective patterns (Fig.~\ref{fig2}). The localization of each individual pattern is exponential; at the distance from the pattern centre $x_0$ the field $\psi\propto\exp(-\gamma|x-x_0|)$, where $\gamma$ is the localization exponent. In~\cite{Goldobin-Shklyaeva-2009,Goldobin-2010} the localization properties of these patterns have been shown to drastically influence the diffusive transport of a passive scalar in the system.

Noise realizations are visually represented in plots with
\[
q_l(x)\equiv
\frac{1}{l}\int_{x-l/2}^{x+l/2}q(x_1)\,\mathrm{d}x_1\,.
\]
When at the point $x$ the value of $q_l(x)$ is positive for large enough $l$ ($l\sim 1$), one may expect a convective flow probably to arise in the vicinity of this point, while in the domain of negative $q_l$ the convective flow is averagely damped. Naturally, as the magnitude of negative $q_0$ increases for a given realization $\xi(x)$, the localized convective patterns become more rare in space.

In~\cite{Goldobin-Shklyaeva-2013}, advection $u$ has been revealed to have a significant influence on the localization properties of patterns. However, the effect of advection was considered there only for small values of $u$, while the strong advection must also suppress the localized convective patterns by washing-out the temperature perturbation from the excitation zone. In this paper we study possible bifurcation scenarios of such a washing-out. Specifically, we firstly argue that the temporal dynamics of the system under consideration is infinitely smooth. Secondly, we provide observations on the possible scenarios with extensive numerical simulations. Thirdly, we derive null-dimensional model reduction equations admitting analytical treatment and find the picture of possible bifurcation scenarios, which is identical to the one reported with numerics.

\section{Two scenarios of washing-out}
Firstly, let us infer what kind of behaviour can be expected as the advection strength $u$ in system~(\ref{eq2-01}) increases. On the one hand, for $u=0$ all patterns are time-independent, which should be preserved by continuity at least for small but finite $u$. On the other hand, for $u>0$ and homogeneous $q(x)=q_0$, due to the translational symmetry, one can eliminate the $u$-term from Eq.~(\ref{eq2-01}) by choosing the coordinate frame moving with velocity $u$. Hence, for a homogeneous $q(x)=q_0$ and nonzero $u$, one should observe space-periodic patterns moving with velocity $u$, which means oscillations at a given point $x$. The frozen noise breaks the space-shift symmetry and, moreover, for the conditions of excitation of localized convective patterns the patterns are pinned to their excitation centres. Thus, for small $u$ the frozen disorder enforces the oscillating patterns to become time-independent. Nonetheless, for some conditions one can expect to observe oscillatory patterns, while the existence of the conditions, for which a convective pattern remains time-independent until it is completely suppressed by advection $u$, is not {\it a priori} excluded. Thus, we can generally expect at least two scenarios: (i)~a scenario with only time-independent patterns and (ii)~scenario with bifurcation involving appearance of stable oscillatory regimes.

\subsection{Smoothness of temporal evolution of system~(\ref{eq2-01})}
\label{sec21}
If we expect existence of stable time-dependent patterns, some general explanations about the temporal evolution of system~(\ref{eq2-01}) with frozen $\delta$-correlated noise $\xi(x)$ are due. Indeed, $\delta$-correlated noise in Eq.~(\ref{eq2-01}) makes $\theta(x,t)$ differentiable with respect to $x$ only twice, as $\theta_{xxx}\sim \sigma\xi(x)\theta_x$; therefore, one can also raise the question of differentiability of $\theta(x,t)$ with respect to time.

Let us consider smoothness of the temporal evolution of Eq.~(\ref{eq2-01}). Nonsmoothness is associated with the highest-order derivatives with respect to $x$ and noise in $q(x)$ and the impact of the advection is of interest---since for $u=0$ only time-independent patterns are stable. Meanwhile, the nonlinear term makes a trivial influence on the dynamics of system~(\ref{eq2-01}) by preventing an infinite growth of excited patterns. Hence, addressing the issue of the smoothness of temporal evolution of $\theta(x,t)$, we can omit the nonlinear term and consider linear equation
\begin{equation}
\dot{\theta}=\left(-u\theta-\theta_{xxx}-q(x)\,\theta_x\right)_x\,.
\label{eq3-01}
\end{equation}
One can consider an eigen-mode of the latter equation $\theta=\Theta(x)\,e^{\lambda t}$. Similarly to the classical Anderson localization, in our case every eigen-mode $\Theta(x)$ is exponentially localized in space~\cite{Goldobin-Shklyaeva-2013} (see Fig.~\ref{fig2} for illustration as well). For the eigen-mode, Eq.~(\ref{eq3-01}) reads
\[
\lambda\Theta=-u\,\Theta_x-\Theta_{xxxx}-\left(q(x)\,\Theta_x\right)_x\,.
\]
Multiplying the latter equation by $\Theta^\ast(x)$ and integrating over $x$, performing similar manipulations for its complex-conjugate and employing partial integration, one can obtain
\begin{equation}
\mathrm{Re}(\lambda)\left\|\,|\Theta|^2\right\|=
 -\left\|\,|\Theta_{xx}|^2\right\|+\left\|\,{q(x)\,|\Theta_x|^2}\right\|\,,
\label{eq3-02}
\end{equation}
\begin{equation}
\mathrm{Im}(\lambda)\left\|\,|\Theta|^2\right\|=
 -2u\big\|\mathrm{Re}(\Theta)\,\mathrm{Im}(\Theta)_x\big\|\,,
\label{eq3-03}
\end{equation}
where $\|...\|\equiv\int_{-\infty}^{+\infty}...\mathrm{d}x$.
One can rewrite equation~(\ref{eq3-02}) as
\begin{eqnarray}
\mathrm{Re}(\lambda)\left\|\,|\Theta|^2\right\|=
 -\left\|\,|\Theta_{xx}|^2\right\|+q_0\left\|\,|\Theta_x|^2\right\|\qquad
\nonumber
\\[5pt]
 {}-\sigma\left\|\,W(x)\,(|\Theta_x|^2)_x\right\|\,,
\label{eq3-04}
\end{eqnarray}
where $W(x)=\sigma^{-1}\!\int^x\xi(x_1)\,\mathrm{d}x_1$ is a Wiener process, which is a finite random number for finite $x$. If $\Theta_x(x)$ is differentiable, {\it i.e.}, $\Theta_{xx}(x)$ is finite, then $(|\Theta_x|^2)_x$ is finite and thus all terms in the r.h.s.\ of (\ref{eq3-04}) are finite; therefore, $\mathrm{Re}(\lambda)$ is finite. Simultaneously, according to~(\ref{eq3-03}), $\mathrm{Im}(\lambda)$ should be finite. If we now admit $\Theta_{xx}(x)$ to tend to infinity, then $\|\,|\Theta_{xx}|^2\|$ diverges as the square of $\Theta_{xx}$, while $\|\,W(x)\,(|\Theta_x|^2)_x\|$ diverges linearly in $\Theta_{xx}$, which---due to positivity of $\|\,|\Theta_{xx}|^2\|$ and $\|\,|\Theta|^2\|$---means that $\mathrm{Re}(\lambda)\to-\infty$. Thus, the eigen-modes with diverging $\Theta_{xx}(x)$ decay infinitely fast in time. To conclude, the eigen-modes either evolve with finite values of $\lambda$ or, with diverging $\Theta_{xx}(x)$, decay infinitely fast and do not contribute to the system dynamics at finite time scales. As any initial state can be decomposed into eigen-modes, the temporal evolution of the system must be infinitely smooth after a vanishingly short transient period.

This property of smoothness of the dependence on one coordinate (time for~(\ref{eq3-01})) in spite of the disorder and nonsmoothness in the dependence on the other coordinate (spatial coordinate $x$ for~(\ref{eq3-01})) is not pertaining uniquely to the system we consider. One can recall a ``mirror'' situation for the dynamics of an ensemble of oscillators with identical or slightly non-identical parameters, subject to identical common noise $\xi(t)$. For such ensembles in~\cite{Yu-Ott-Chen-1990,Yu-Ott-Chen-1991}, the notion of a snapshot attractor---the ensemble state at certain instant of time---has been introduced. In spite of random non-smooth dynamics of oscillators in time, the snapshot attractors, which represent a kind of a spatial pattern, turn out to be smooth and regular (can be fractal, but still regular with a fine structure)~\cite{Yu-Ott-Chen-1990,Yu-Ott-Chen-1991,Goldobin-Pikovsky-2005,Goldobin-2008}. Here, the situation is similar to the case of Eq.~(\ref{eq2-01}) up to the interchange of the space and time coordinates.


\subsection{Bifurcation scenarios}
\begin{figure}[!t]
\center{
(a)
\includegraphics[width=0.43\textwidth]%
{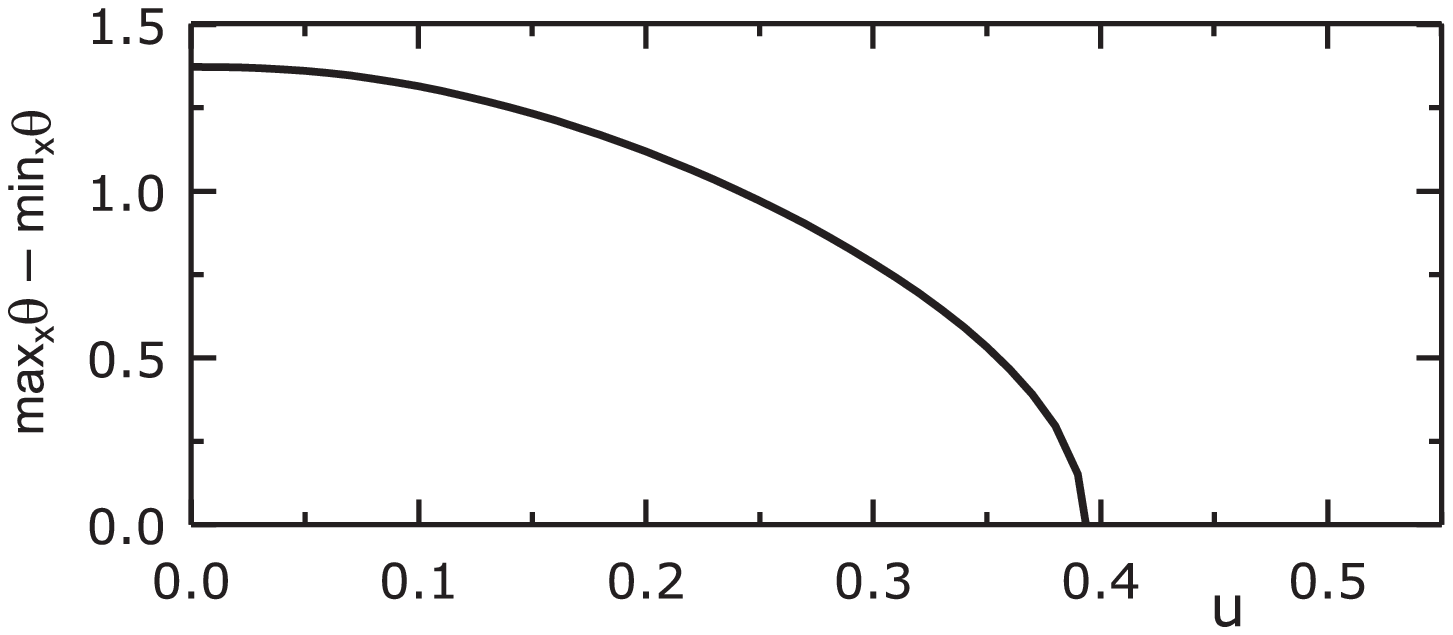}\\[5pt]
(b)
\includegraphics[width=0.43\textwidth]%
{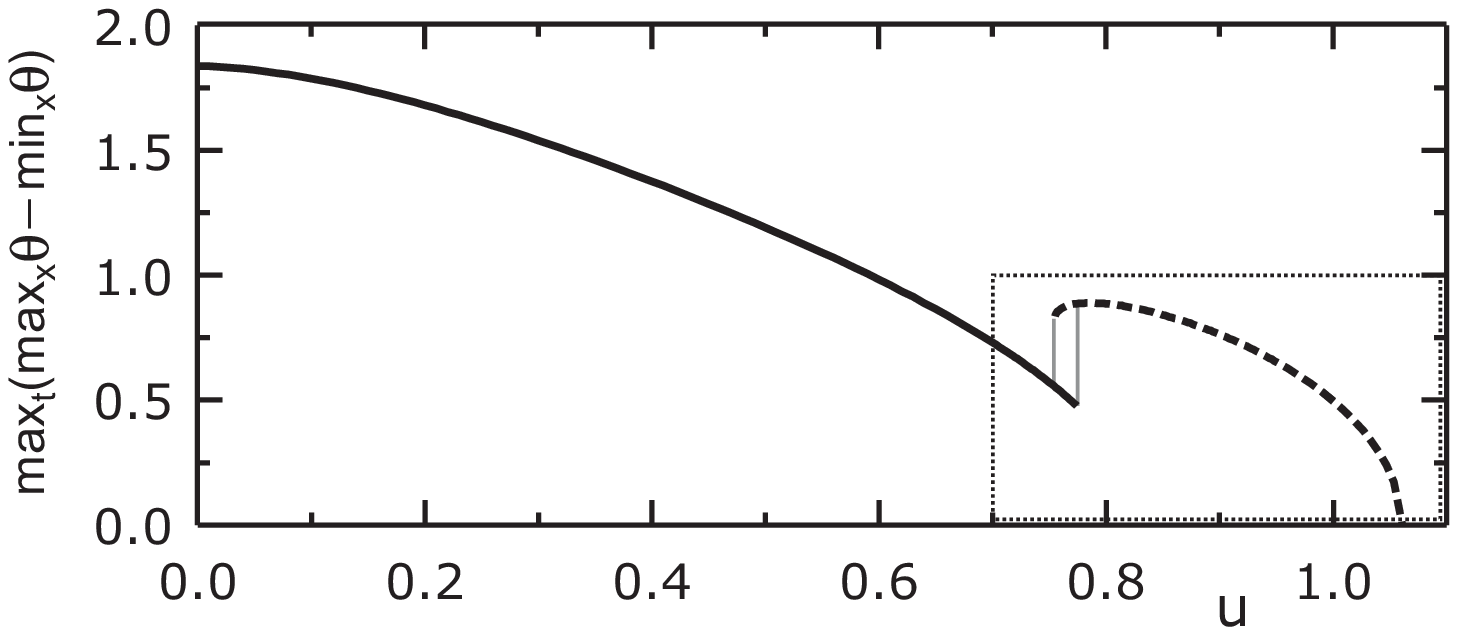}\\[5pt]
(c)
\includegraphics[width=0.43\textwidth]%
{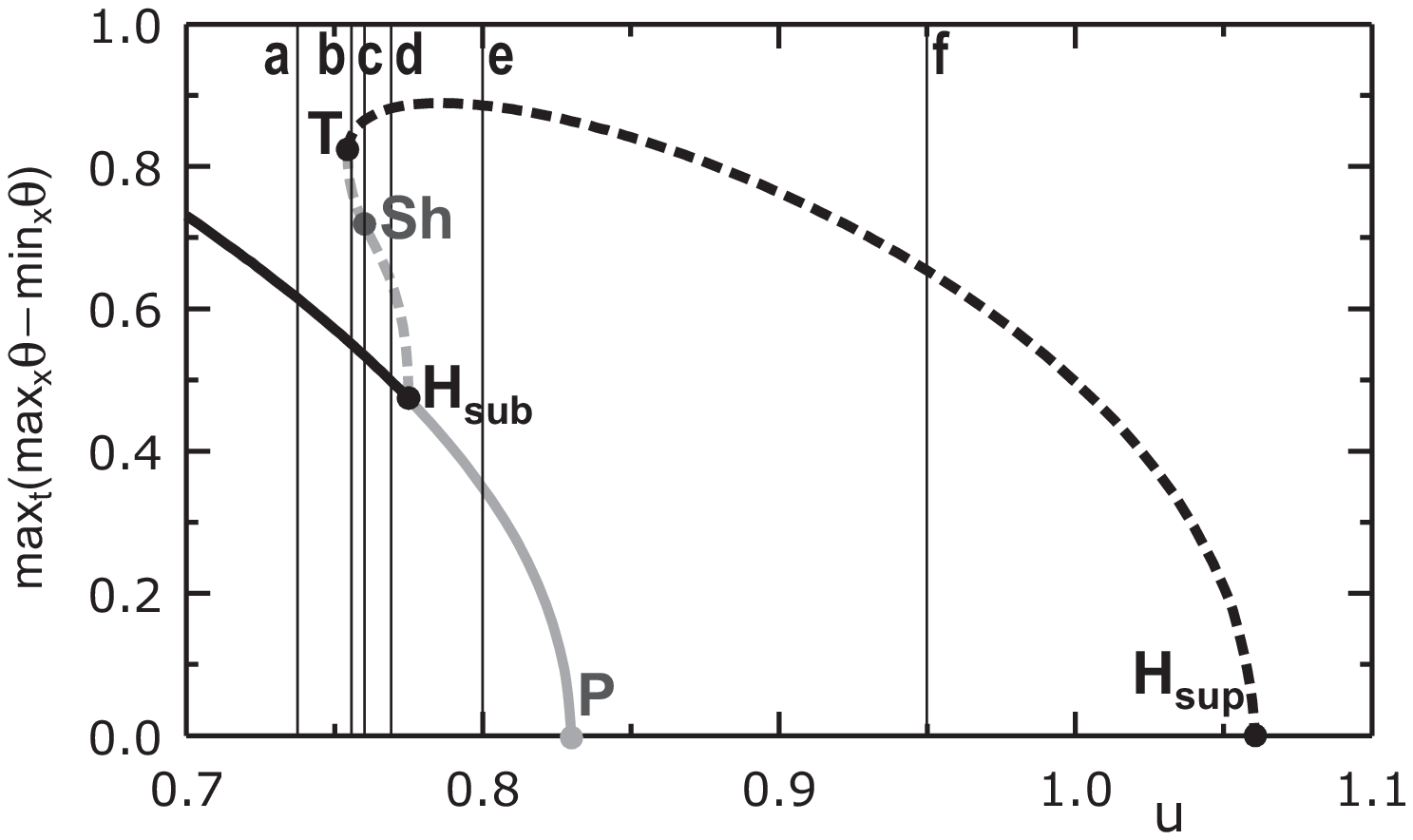}\\[5pt]
(d)
\includegraphics[width=0.43\textwidth]%
{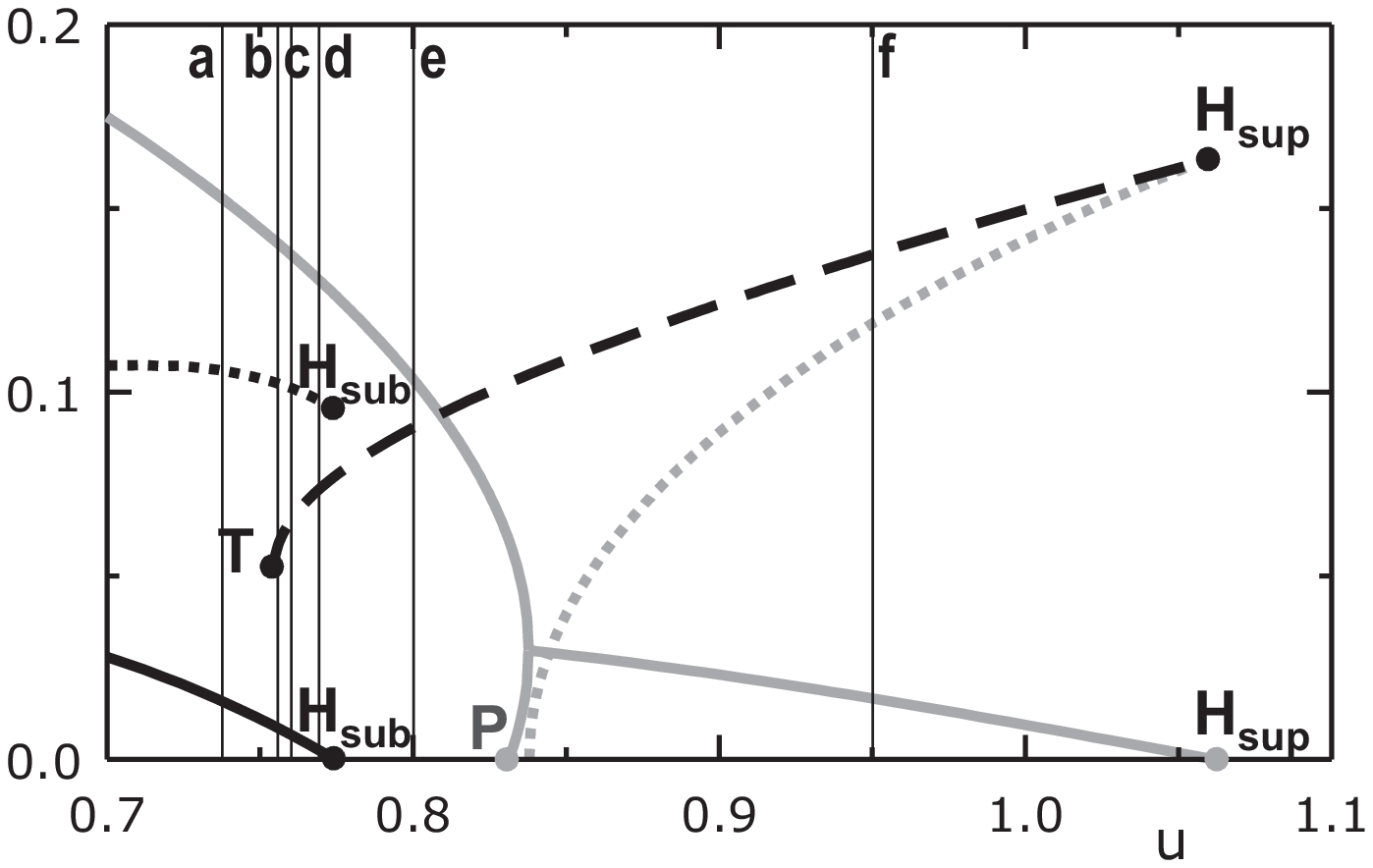}\\
}		
\caption{{\bf Scenario~I}:
(a)~The maximal temperature difference of a localized pattern for a sample realization of noise is plotted {\it vs} advection strength $u$. The pattern is time-independent for all $u$.
{\bf Scenario~II}:
(b)~The maximal temperature difference $[\theta]$ of the time-independent stable pattern (solid line) and the maximal value of $[\theta]$ for the stable oscillatory regime (dashed line) are plotted {\it vs} advection strength $u$ for a sample realization of $q(x)$ presented in Fig.~\ref{fig5}a.
(c)~Zoom-in of the region marked in plot~(b), gray lines show unstable regimes.
(d)~Black lines: the exponential decay rate (solid line) and the frequency (dotted line) of perturbations of the time-independent pattern and the frequency of the stable oscillatory flow (dashed line), gray lines: the exponential growth rate (solid) and the frequency (dotted) of perturbations of the trivial state.
For the values of $u$ marked by vertical lines with letters in plots~(c,d), projections of the phase portraits are shown in Fig.~\ref{fig4} under the corresponding letters.
}
\label{fig3}
\end{figure}

In numerical simulations (for technical details on the finite difference scheme and its validation see Appendix in~\cite{Goldobin-Shklyaeva-2013}) we observed only two bifurcation scenarios of advective suppression of localized convective patterns. Scenario~I is trivial; the stable patterns remain time-independent for all values of $u$, the pattern amplitude monotonously decreases with increasing $u$ and at certain value $u_\mathrm{cr}$ the stable convective patterns disappear via a pitchfork bifurcation (Fig.~\ref{fig3}a). The bifurcation is a pitchfork one due to the symmetry $\theta\leftrightarrow-\theta$.

Let us now consider a complex scenario~II for a sample realization of $q(x)$ shown in Fig.\,\ref{fig5}a and periodic lateral boundary conditions. In Figs.\,\ref{fig3}b-d, the maximal temperature difference $[\theta]$ of the stable convective patterns is plotted {\it vs} advection strength $u$ and the stability properties of these patterns and the trivial state are presented. In Fig.~\ref{fig3}b, one can see that the time-independent solution, which exists at $u=0$, becomes unstable for advection $u$ strong enough. Meanwhile, some stable oscillatory regime emerges in the system. There is a hysteresis of the transition between these two regimes. With further increase of the advection strength $u$ the oscillation amplitude decreases, until the oscillatory regime disappears via a Hopf bifurcation at $u_\mathrm{cr}$; $u_\mathrm{cr}$ is the critical strength of advection required for suppression of the localized convective pattern.

\begin{figure}[!t]
\center{\includegraphics[width=0.480\textwidth]%
{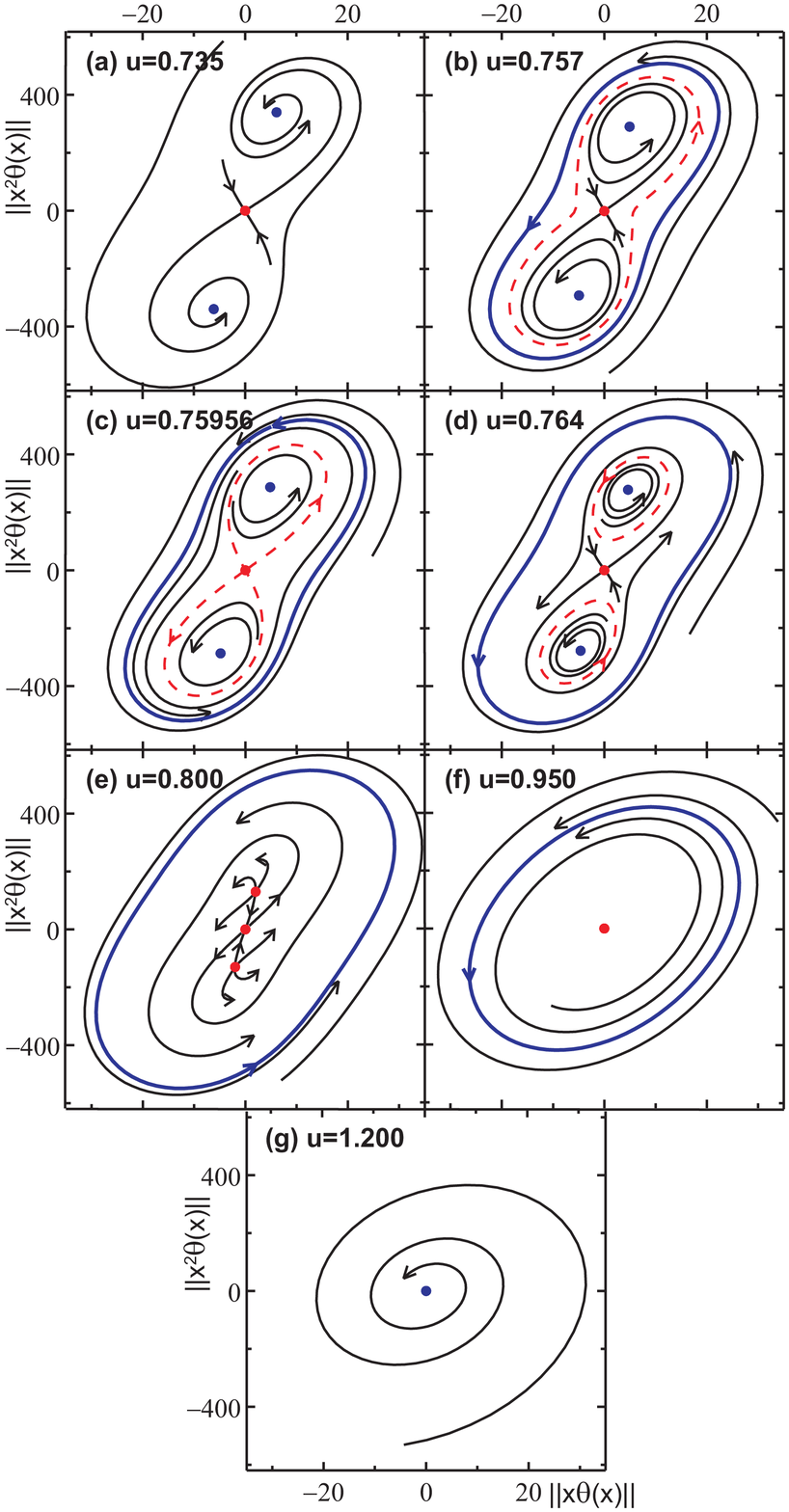}}

\caption{
Cascade of the projections of the phase trajectories of system~(\ref{eq2-01}) onto the plane $(\|x\theta\|,\|x^2\theta\|)$ for $q(x)$ plotted in Fig.~\ref{fig5}a and specified values of advective through-flow $u$.}
\label{fig4}
\end{figure}

\begin{figure*}[!t]
\center{
\begin{tabular}{ccc}
(a)\hspace{-5mm}\includegraphics[width=0.42\textwidth]%
{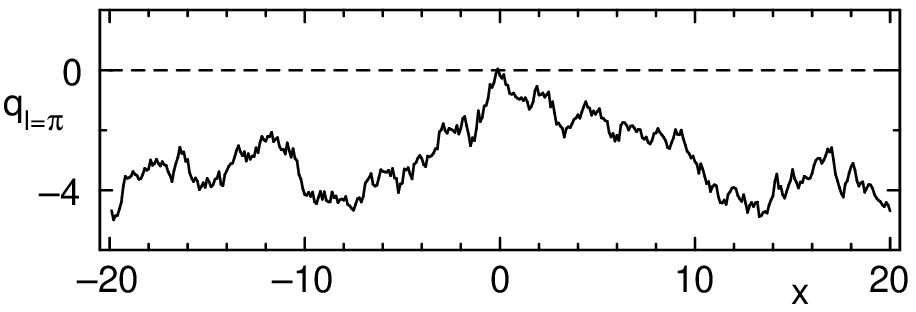}
&\qquad&
(b)\hspace{-5mm}\includegraphics[width=0.42\textwidth]%
{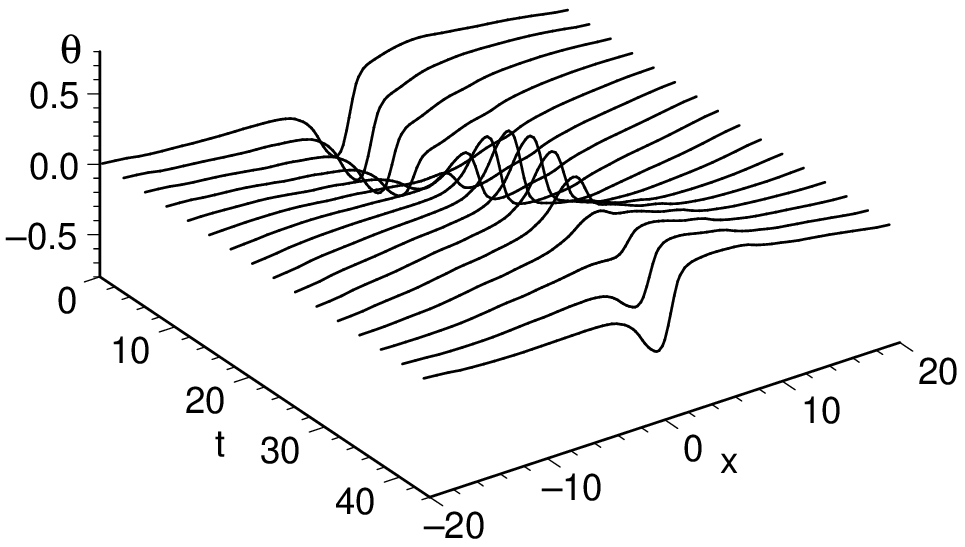}\\[20pt]
(c)\hspace{-5mm}\includegraphics[width=0.42\textwidth]%
{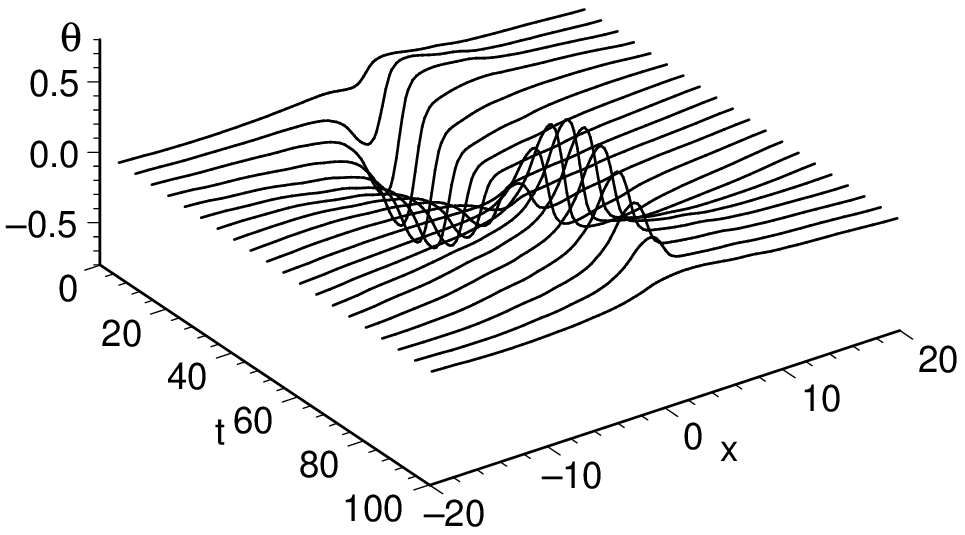} & &
(d)\hspace{-5mm}\includegraphics[width=0.42\textwidth]%
{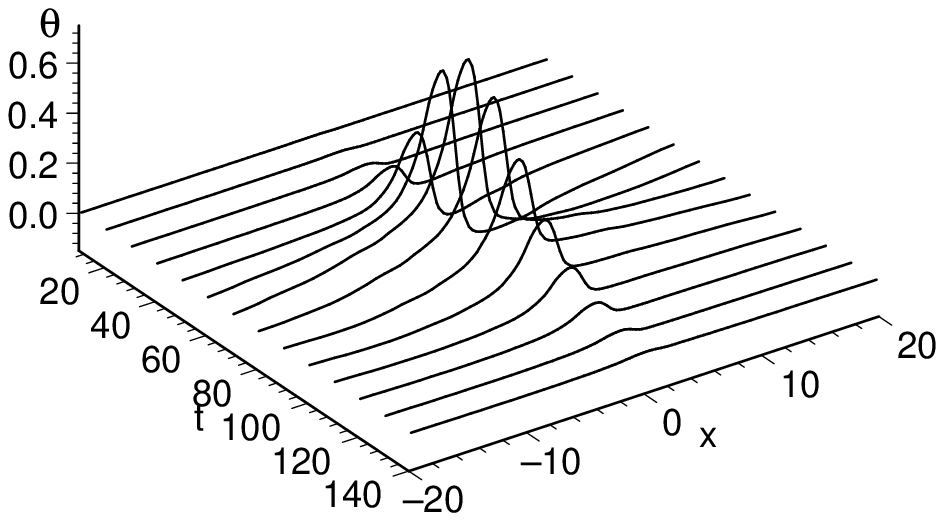}
\end{tabular}
}

\caption{
Dynamics of field $\theta(x,t)$ is presented for the sample realization of $q(x)$ plotted in (a).
(b):~the stable time-periodic pattern at $u=0.95$,
(c,d):~the stable time-periodic pattern and the homoclinic pattern at $u_\mathrm{Sh}\approx0.75956$, respectively.}
\label{fig5}
\end{figure*}

For a comprehensive understanding of this bifurcation scenario, we provide a zoom-in of the amplitude diagram of the observed regimes in Fig.~\ref{fig3}c and the linear stability analysis of the observed regimes in Fig.~\ref{fig3}d.
Close to point {\bf P}
the exponential growth rate of the linear perturbations of the trivial state is small and turns complex (gray curves in Fig.~\ref{fig3}d). Hence, the system is close to a co-dimension 2 bifurcation where two real eigenvalues of the linear stability problem turn to zero, but this bifurcation differs from the generic Bogdanov--Takens one~\cite{Kuznetsov-2004} due to the sign inversion symmetry ($\theta\to-\theta$) of our system. Such a situation (being close to a Bogdanov--Takens bifurcation modified by the sign inversion symmetry) is widespread in fluid dynamics; {\it e.g.}, Knobloch and Proctor have encountered it for the problem of convection in the presence of an imposed vertical magnetic field~\cite{Knobloch-Proctor-1981}.
On some stages of the bifurcation scenario the principal manifold of the system is two-dimensional (cf~\cite{Knobloch-Proctor-1981}) and, therefore, the minimal dimension required for capturing the system dynamics is 2. We project the system trajectories onto the plane $(\|x\theta\|,\|x^2\theta\|)$ for presentation in Fig.~\ref{fig4} (here $\|...\|\equiv\int_{-L_\mathrm{layer}/2}^{L_\mathrm{layer}/2}...\,\mathrm{d}x$ stands for the integration over the calculation domain of length $L_\mathrm{layer}$); linear in $\theta$ quantifiers $\|x\theta\|$ and $\|x^2\theta\|$ are the simplest independent integral quantifiers of pattern $\theta(x)$.

In the absence of advection, only time-independent convective patterns can arise in dynamical system~(\ref{eq2-01}). For small $u$, this holds true by continuity (Fig.~\ref{fig4}a). As advection becomes stronger, a pair of finite-amplitude time-periodic regimes (Fig.~\ref{fig4}b) appears at $u=u_\mathrm{T}\approx0.754$ via a tangent bifurcation (point {\bf T} in Figs.~\ref{fig3}c,d); the more intense periodic pattern is locally stable (Fig.~\ref{fig4}b, solid loop: stable limit cycle, dashed curve: saddle cycle). At $u_\mathrm{Sh}\approx0.75956$ ({\bf Sh} in Fig.~\ref{fig3}c), the unstable (metastable) manifold of the trivial state intersects the stable one (more specifically, the one-dimensional unstable manifold is embedded in the hypersurface of the stable one), and two homoclinic loops form (see Fig.~\ref{fig4}c, in Fig.~\ref{fig5}d the temporal evolution of the homoclinic solution is presented); the union of these loops is the former unstable cycle. In Fig.~\ref{fig4}c, simultaneously with these repelling loops, the locally stable periodic (Fig.~\ref{fig5}c) and time-independent patterns can bee seen; the homoclinic loops belong to the boundary between the attraction basins of these stable regimes---in the two-dimensional setup of Fig.~\ref{fig4} the one-dimensional loops visually exhaust the boundary which is a hypersurface of co-dimension 1. For stronger advection, $u_\mathrm{Sh}<u<u_\mathrm{H_{sub}}$, the homoclinic loops turn into a pair of unstable cycles (Fig.~\ref{fig4}d). This bifurcation of homoclinic trajectories is called a gluing bifurcation, which somewhat differs from a generic bifurcation described by the Shil'nikov's theorem~\cite{Shilnikov-1966,Kuznetsov-2004} due to the symmetry $\theta\to-\theta$. To summarize, as a result of this bifurcation, the big unstable periodic orbit with three equilibria (a saddle and two foci) inside it is transformed into two smaller orbits, each one around a single equilibrium. Further, the unstable cycles shrink as $u$ increases and collapse onto the time-independent solutions making them unstable via a subcritical Hopf bifurcations at $u_\mathrm{H_{sub}}\approx0.775$ ($\mathrm{\bf H_{sub}}$ in Figs.~\ref{fig3}c,d). For $u_\mathrm{H_{sub}}<u<u_\mathrm{H_{sup}}$, the only stable regime is the time-periodic pattern (Figs.~\ref{fig4}e,f, \ref{fig5}b). Within this range of $u$, at $u_\mathrm{P}\approx0.832$, the unstable time-independent patterns disappear via a pitchfork bifurcation near the trivial state ({\bf P} in Figs.~\ref{fig3}c,d), turning it from a saddle into an unstable node, which quickly becomes a focus after further increase of $u$ (Fig.~\ref{fig4}f). For $u>u_\mathrm{P}$, increase of $u$ leads to the shrinking of the limit cycle till the time-periodic pattern diminishes to zero at $u_\mathrm{H_{sup}}\approx1.062$ and disappears via a supercritical Hopf bifurcation ($\mathrm{\bf H_{sup}}$ in Figs.~\ref{fig3}c,d). For $u>u_\mathrm{H_{sup}}$, all convective currents decay (Fig.~\ref{fig4}g). Noticeably, the vicinity of trivial state belongs to the attraction basin of the time-independent patterns for $u<u_\mathrm{Sh}$ and to the one of the time-periodic pattern for $u_\mathrm{Sh}<u<u_\mathrm{H_{sup}}$, which determines whether small initial perturbations of the no-flow state evolve to the former regime or to the latter.

\begin{figure*}[!t]
\center{
\begin{tabular}{ccc}
\includegraphics[width=0.315\textwidth]%
{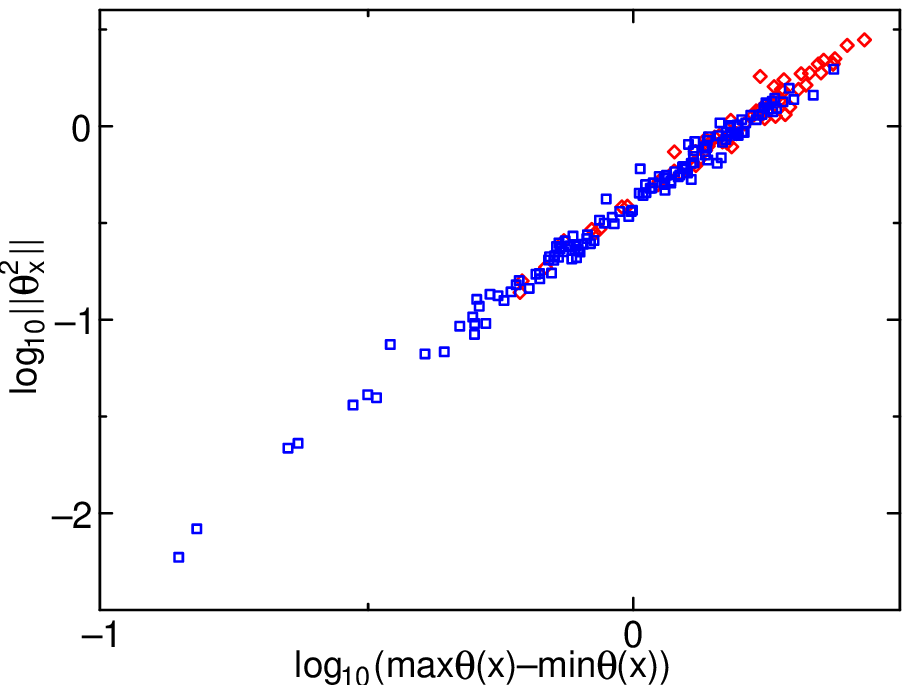}
&
\includegraphics[width=0.315\textwidth]%
{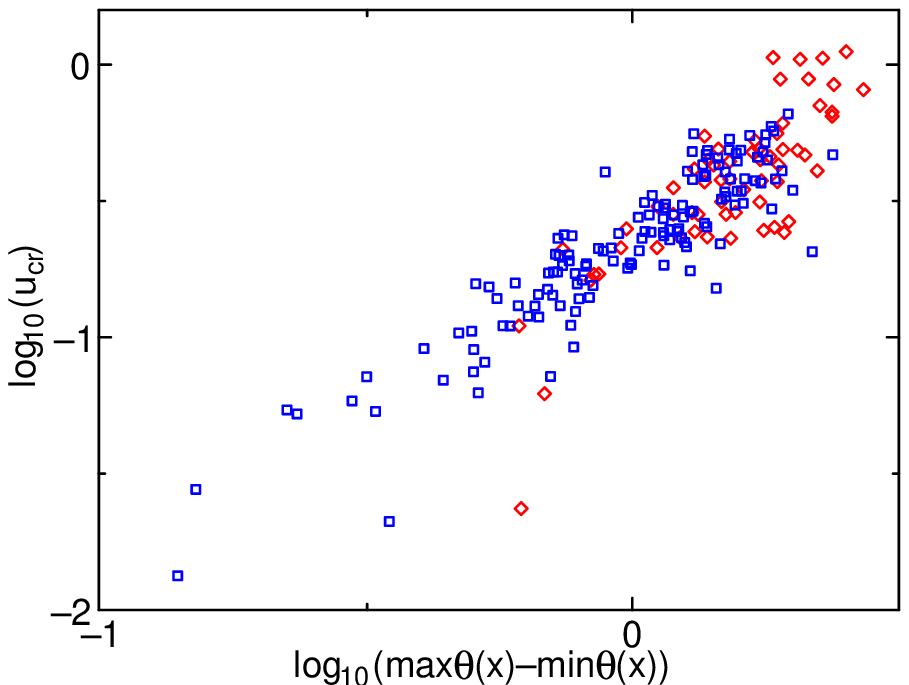}
&
\includegraphics[width=0.315\textwidth]%
{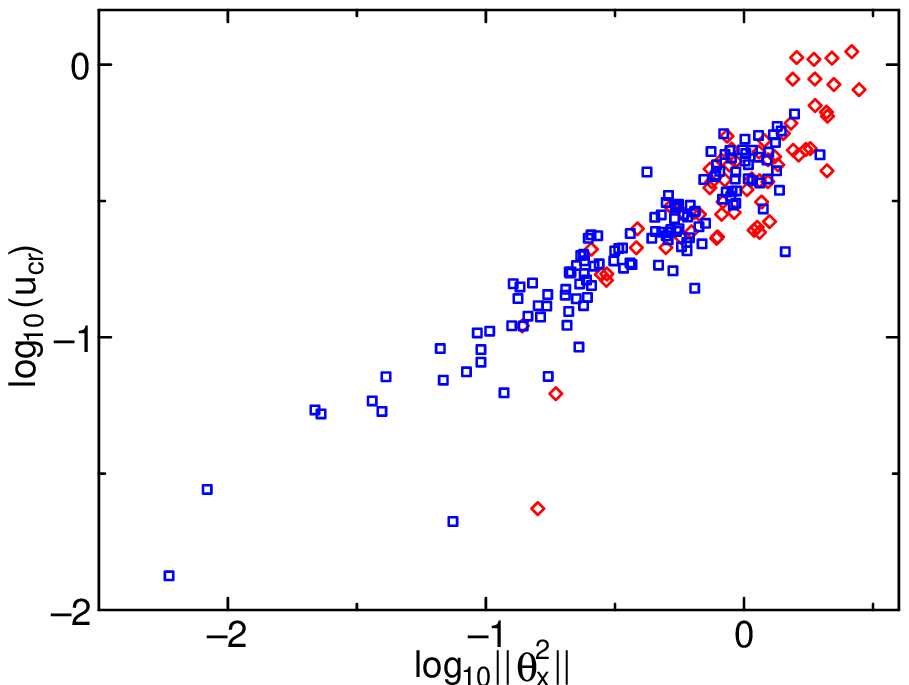}\\[5pt]
(a) & (b) & (c)
\end{tabular}
}
\caption{
Integral characteristics of localized patterns at $q_0=-3$, $u=0$. (a):~flow intensity $\|\theta_x^2\|$ {\it vs} the maximal temperature difference $[\theta]$; (b) and (c):~critical advection strength $u_\mathrm{cr}$ {\it vs} $[\theta]$ and $\|\theta_x^2\|$, respectively. Blue squares: the convective pattern disappears via a pitchfork bifurcation, red diamonds: via a Hopf bifurcation.
}
\label{fig6}
\end{figure*}

The difference in the scenarios of advective suppression for two groups of localized patterns raises the natural question whether these patterns can be distinguished at $u=0$ and what (if any) can be an appropriate discrimination characteristic.
The following numerical simulation was performed in order to address this question.
Firstly, long records of $q(x)$ (of length $10\,000$) with properties~(\ref{eq2-02}) were randomly generated. Secondly, at $u=0$, with the algorithm technically described in Appendix of Ref.~\cite{Goldobin-Shklyaeva-2013}, all pattern excitation centres for the linearized equation~(\ref{eq3-01}) were identified and cut-out with the vicinity of length $40$. Thirdly, at $u=0$, the nonlinear equation~(\ref{eq2-01}) was numerically integrated in each cut-out domain of $q(x)$ until the formation of a time-independent state; for each excitation centre, the normalized growing eigenfunction of the linearized problem was used as the initial condition. Fourthly, $u$ was gradually increased until the decay of steady regimes (for $u\ne0$ they can be both time-independent and oscillatory) to zero at $u_\mathrm{cr}$.

In Fig.~\ref{fig6}, the integral characteristics of localized patterns observed in numerical simulations are plotted for $q_0=-3$.
In Fig.~\ref{fig6}a, one can see, that for $u=0$ the states swarm around the trend $\|\theta_x^2\|=k[\theta]^2$. The proportionality coefficient $k$, as well as the dispersion of states around the trend, are similar for both the patterns disappearing via a pitchfork bifurcation (scenario~I) and a Hopf one (scenario~II), meaning the patterns of these two types can be hardly distinguished by their simplest shape properties.
In Figs.~\ref{fig6}b and c, one can see that critical values $u_\mathrm{cr}$ for a given value of $[\theta]$ or $\|\theta_x^2\|$ are strongly dispersed---almost by an order of magnitude. Thus, the primary quantifiers of the pattern intensity are very far from being sufficient for prediction of the advection strength required for the pattern suppression. The only reliable conclusion which can be drawn from the numerical results plotted in Fig.~\ref{fig6} is that intense patterns more frequently disappear via a Hopf bifurcation, while for the low-intensity patterns a pitchfork bifurcation is more typical. However, this qualitative discrepancy is quite weak; the majorities of both scenarios are observed in the range of intensities where they coexist. To conclude, the correlation between scenarios of advective suppression and integral quantifiers of the convective pattern intensity turns out to be not playing a decisive role.

\section{Two-mode model reduction}
For better understanding of these scenarios and interrelations between them, let us analyse the model reduction admitting comprehensive consideration.

\subsection{Derivation of model reduction equations}
Let us consider a localized pattern in system~(\ref{eq2-01}). This localized pattern onsets where the ground state of the system is unstable. Mathematically it means that the linearized equation (\ref{eq3-01})  possesses a spatially localized instability mode. The localized solutions are of interest when they are sparse in space, which is the case of non-small negative $q_0$ (see Fig.~\ref{fig2} and~\cite{Goldobin-Shklyaeva-2013}); therefore, one can rarely observe more than one instability mode at certain excitation centre. Thus, we rewrite Eq.~(\ref{eq2-01}) as
\begin{equation}
\dot\theta=-u\theta_x+\hat{L}\theta+\big((\theta_x)^3\big)_x\,,
\label{eq4-01}
\end{equation}
where
\begin{equation}
\hat{L}f\equiv-f_{xxxx}-\big(q(x)\,f_x\big)_x\,,
\label{eq4-02}
\end{equation}
and assume only one instability mode at $u=0$\,;
\begin{equation}
\hat{L}\,\theta_\lambda(x)=\lambda\,\theta_\lambda(x)
\label{eq4-03}
\end{equation}
with $\lambda>0$. In Sec.~\ref{sec21}, we demonstrated that the exponential growth rate $\lambda$ is real-valued at $u=0$ (in Eq.~(\ref{eq3-03}), $\mathrm{Im}(\lambda)\propto u$).

The nonlinear term in Eq.~(\ref{eq4-01}) prevents the instability mode from infinite growth, resulting in saturation, while advection $u$ affects the shape of the pattern. Specifically, the $u$-term makes contribution $-u\theta_x$ into the time-derivative $\dot\theta$. Hence, for non-large $u$ and $\lambda$, the shape of the temperature field is contributed by the principal part $\propto\theta_\lambda(x)$ and perturbed by $\theta_x$\,; a sound approximation for the temperature field becomes
\begin{equation}
\theta(x,t)\approx a(t)\,\theta_\lambda(x)+b(t)\,\theta_\lambda^\prime(x)\,,
\label{eq4-04}
\end{equation}
(henceforth, the prime stands for the $x$-derivative).
Eq.~(\ref{eq4-01}) with temperature field (\ref{eq4-04}) can be projected onto $\theta_\lambda(x)$ and $\theta_\lambda^\prime(x)$ with a standard definition of the inner product;
one finds
\begin{eqnarray}
&&\dot{a}\left\|\,\theta_\lambda^2\right\|=
ub\left\|\,(\theta_\lambda^\prime)^2\right\|
 +\lambda a\left\|\,\theta_\lambda^2\right\|
 -a^3\left\|\,(\theta_\lambda^\prime)^4\right\|
\nonumber\\[5pt]
&&\qquad{}
 -3ab^2\left\|\,(\theta_\lambda^\prime)^2(\theta_\lambda^{\prime\prime})^2\right\|
 -b^3\left\|\,\theta_\lambda^\prime\,(\theta_\lambda^{\prime\prime})^3\right\|\,,
\label{eq4-05}\\[5pt]
&&\dot{b}\left\|\,(\theta_\lambda^\prime)^2\right\|=
 -ua\left\|\,(\theta_\lambda^\prime)^2\right\|
 +b\left\|\,\theta_\lambda^\prime\hat{L}\theta_\lambda^\prime\right\|
\nonumber\\[5pt]
&&\qquad{}
 -3a^2b\left\|\,(\theta_\lambda^\prime)^2(\theta_\lambda^{\prime\prime})^2\right\|
\nonumber\\[5pt]
&&\qquad\qquad{}
 -3ab^2\left\|\,\theta_\lambda^\prime\,(\theta_\lambda^{\prime\prime})^3\right\|
 -b^3\left\|\,(\theta_\lambda^{\prime\prime})^4\right\|\,.
\label{eq4-06}
\end{eqnarray}
Here partial integration has been employed to evaluate the following integrals:
\begin{eqnarray}
&&
\left\|\,\theta_\lambda\theta_\lambda^\prime\right\| =\left\|\,(\theta_\lambda^2/2)^\prime\right\| =(\theta_\lambda^2/2)\big|_{-\infty}^{+\infty}=0\,,
\nonumber\\
&&
\left\|\,\theta_\lambda\theta_\lambda^{\prime\prime}\right\| =-\left\|\,(\theta_\lambda^\prime)^2\right\|\,,
\nonumber\\
&&
\left\|\,\theta_\lambda\hat{L}\theta_\lambda^\prime\right\| =\left\|\,(\hat{L}\theta_\lambda)\,\theta_\lambda^\prime\right\| =\lambda\left\|\,\theta_\lambda\theta_\lambda^\prime\right\|=0\,,
\nonumber\\
&&
\left\|\,\theta_\lambda\big((\theta^\prime)^3\big)^\prime\right\| =-\left\|\,\theta_\lambda^\prime(\theta^\prime)^3\right\|
\nonumber\\
&&\qquad{}
=-a^3\left\|\,(\theta_\lambda^\prime)^4\right\|
-3a^2b\left\|\,(\theta_\lambda^\prime)^3\theta_\lambda^{\prime\prime}\right\|
\nonumber\\
&&\qquad\qquad{}
-3ab^2\left\|\,(\theta_\lambda^\prime)^2(\theta_\lambda^{\prime\prime})^2\right\|
-b^3\left\|\,\theta_\lambda^\prime\,(\theta_\lambda^{\prime\prime})^3\right\|\,,
\nonumber\\
&&\left\|\,\theta_\lambda^\prime\big((\theta^\prime)^3\big)^\prime\right\| =-\left\|\,\theta_\lambda^{\prime\prime}(\theta^\prime)^3\right\|
\nonumber\\
&&\qquad{}
=-a^3\left\|\,(\theta_\lambda^\prime)^3\theta_\lambda^{\prime\prime}\right\|
-3a^2b\left\|\,(\theta_\lambda^\prime)^2(\theta_\lambda^{\prime\prime})^2\right\|
\nonumber\\
&&\qquad\qquad{}
-3ab^2\left\|\,\theta_\lambda^\prime\,(\theta_\lambda^{\prime\prime})^3\right\| -b^3\left\|\,(\theta_\lambda^{\prime\prime})^4\right\|\,,
\nonumber
\end{eqnarray}
where
$\|\,(\theta_\lambda^\prime)^3\theta_\lambda^{\prime\prime}\| =\|\,\big((\theta_\lambda^\prime)^4/4\big)^\prime\| =(\theta_\lambda^4/4)|_{-\infty}^{+\infty}=0$\,.

Let us notice that the integral $\|\,\theta_\lambda^\prime\,(\theta_\lambda^{\prime\prime})^3\|$ turns to zero for both even and odd functions $\theta_\lambda(x)$, meaning that this integral represents a symmetry defect of the mode, while all the other coefficients of nonlinear terms have non-negative integrands. Hence, one can typically expect integral $\|\,\theta_\lambda^\prime\,(\theta_\lambda^{\prime\prime})^3\|$ to be small compared to other integrals. Indeed, specifically for the noise realization presented in Fig.~\ref{fig2} for $q_0=-2.5$, the numerical calculation of integrals yields $\|\,(\theta_\lambda^{\prime})^4\|\approx0.0682$, $\|\,(\theta_\lambda^\prime)^2(\theta_\lambda^{\prime\prime})^2\|\approx0.0561$, $\|\,\theta_\lambda^\prime\,(\theta_\lambda^{\prime\prime})^3\|\approx-0.0059$, $\|\,(\theta_\lambda^{\prime\prime})^4\|\approx0.1198$; the absolute value of the integral under discussion is by one order of magnitude smaller than the other integrals. For the further analysis we will set this integral to zero for simplicity without loss of generality.

In terms of
\[
\begin{array}{c}
\displaystyle
\tau=\lambda t\,,
\qquad
A=a\sqrt{\frac{\left\|\,(\theta_\lambda^\prime)^4\right\|}
{\lambda\left\|\,\theta_\lambda^2\right\|}}\,,
\\[5pt]
\displaystyle
B=b\sqrt{\frac{\left\|\,(\theta_\lambda^\prime)^4\right\|\,\left\|\,(\theta_\lambda^\prime)^2\right\|}
{\lambda\left\|\,\theta_\lambda^2\right\|^2}}\,,
\qquad
U=\frac{u}{\lambda}\sqrt{\frac{\left\|\,(\theta_\lambda^\prime)^2\right\|}
{\left\|\,\theta_\lambda^2\right\|}}
\end{array}
\]
Eqs.~(\ref{eq4-05})--(\ref{eq4-06}) read
\begin{eqnarray}
&&\dot{A}=
A+UB-A^3
 -3AB^2\frac{\left\|\,(\theta_\lambda^\prime)^2(\theta_\lambda^{\prime\prime})^2\right\| \left\|\,\theta_\lambda^2\right\|}
 {\left\|\,(\theta_\lambda^\prime)^4\right\| \left\|\,(\theta_\lambda^\prime)^2\right\|}\,,
\label{eq4-07}\\[5pt]
&&\dot{B}=
 -UA
 +\frac{\left\|\,\theta_\lambda^\prime\hat{L}\theta_\lambda^\prime\right\|} {\lambda\left\|\,(\theta_\lambda^\prime)^2\right\|}B
 -3A^2B\frac{\left\|\,(\theta_\lambda^\prime)^2(\theta_\lambda^{\prime\prime})^2\right\| \left\|\,\theta_\lambda^2\right\|} {\left\|\,(\theta_\lambda^\prime)^4\right\| \left\|\,(\theta_\lambda^\prime)^2\right\|}
\nonumber\\[5pt]
&&\qquad{}
 -B^3\frac{\left\|\,(\theta_\lambda^{\prime\prime})^4\right\| \left\|\,\theta_\lambda^2\right\|^2} {\left\|\,(\theta_\lambda^\prime)^4\right\| \left\|\,(\theta_\lambda^\prime)^2\right\|^2}\,.
\label{eq4-08}
\end{eqnarray}

For large negative $q_0$, when the pattern excitation centres are rare, the probability of heaving the growth rate $\lambda$ rapidly decreases with the increase of $\lambda$; small values of $\lambda$ are abundant, while large values of $\lambda$ are nearly improbable. Simultaneously, other modes are rapidly decaying and a large decay rate $(-\|\,\theta_\lambda^\prime\hat{L}\theta_\lambda^\prime\|/\|\,(\theta_\lambda^\prime)^2\|)$ of $B$ should be typical. Hence, the coefficient
\[
K\equiv\frac{-\left\|\,\theta_\lambda^\prime\hat{L}\theta_\lambda^\prime\right\|} {\lambda\left\|\,(\theta_\lambda^\prime)^2\right\|}
\]
is typically large. Specifically for the noise realization presented in Fig.~\ref{fig2} for $q_0=-2.5$, numerical calculations yield $\|\,\theta_\lambda^\prime\hat{L}\theta_\lambda^\prime\|/ \|\,(\theta_\lambda^\prime)^2\|\approx-13.95$, which is large negative, and $\lambda\approx0.2407$, which is small positive, as expected.

The nonlinear terms in Eqs.~(\ref{eq4-07})--(\ref{eq4-08}) are purely dissipative. While for the excited mode $A$ the nonlinearity deters the growth, the eigen dynamics of perturbation $B$ is a rapid decay ($K$ is large), which is only facilitated by nonlinear terms. The mode $B$ experiencing a strong dumping is pushed from the zero-state merely by the advection term $UA$. Hence, $B$ should be small compared to $A$ and one can neglect all the nonlinear terms except for the leading one $A^3$.

Finally, the model reduction equations relevant for typical localized patterns can be written-down as
\begin{eqnarray}
\dot{A}&= A+UB-A^3,
\label{eq4-09}\\[5pt]
\dot{B}&= -UA-KB\,.
\label{eq4-10}
\end{eqnarray}

\begin{figure*}[!t]
\center{\includegraphics[width=0.65\textwidth]%
{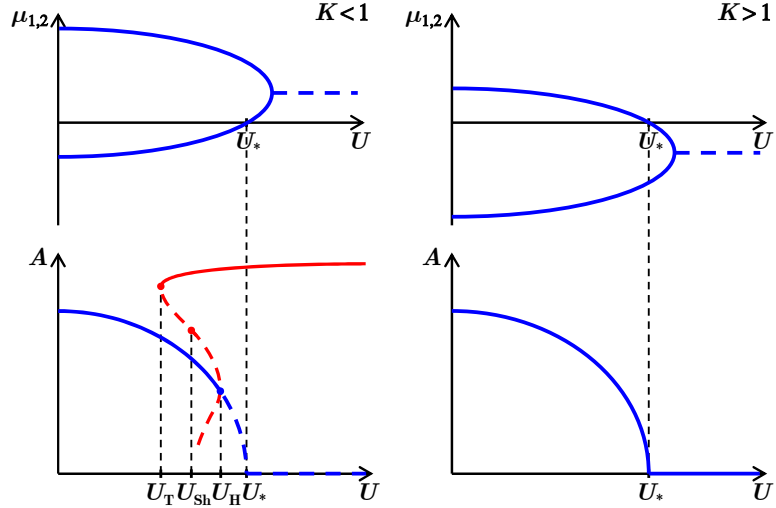}}

\caption{
The dependence of the stability properties of the trivial state of the dynamic system (\ref{eq4-09})--(\ref{eq4-10}) on $U$ is plotted in the upper graphs for $K<1$ and $K>1$. The corresponding finite-amplitude regimes of the system dynamics are presented in the lower graphs. The system dynamics for $K<1$ is shown in further detail in Fig.~\ref{fig8}.}
\label{fig7}
\end{figure*}

\begin{figure*}[!t]
\center{\includegraphics[width=0.95\textwidth]%
{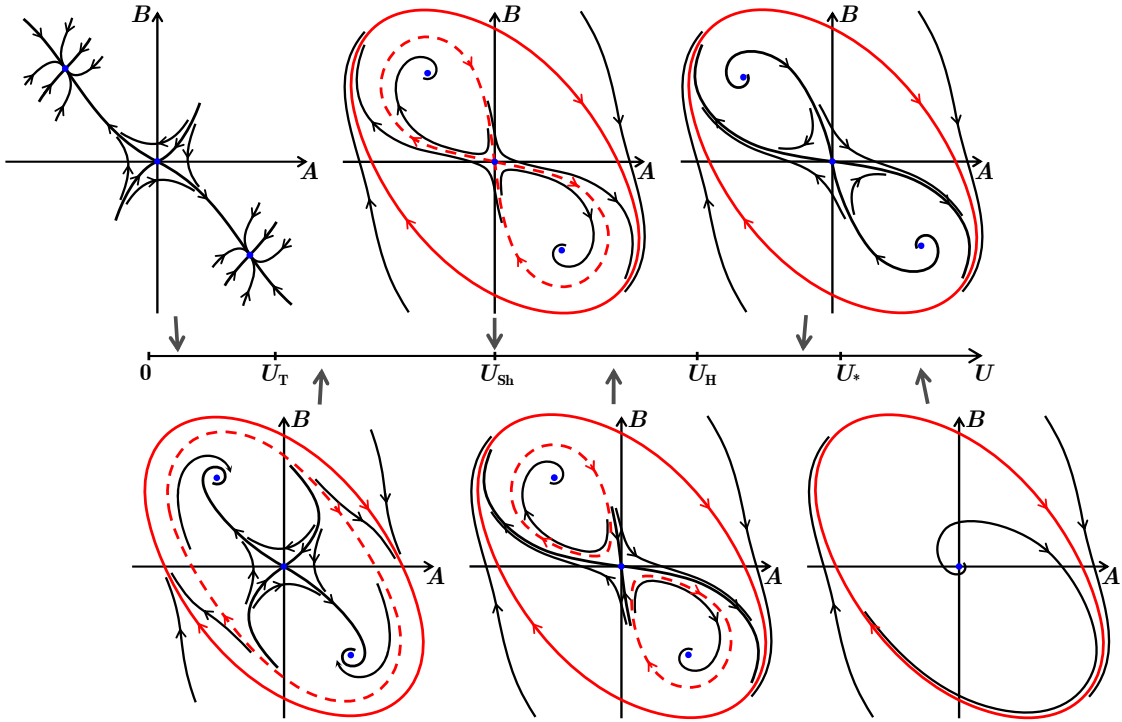}}

\caption{
The bifurcation scenario of the dynamic system (\ref{eq4-09})--(\ref{eq4-10}) is presented for $K<1$. Next to $U=0$, there are three fixed points in the phase space; two nontrivial ones are attracting. At $U_\mathrm{T}$ the pair of stable and unstable cycles arises around these fixed points (the red solid and dashed closed curves, respectively) via a tangent bifurcation. As $U$ increases, the outer stable cycle slightly grows, while the inner unstable one shrinks, until it turns into two homoclinic loops at $U_\mathrm{Sh}$ and further splits into two separate unstable cycles for $U>U_\mathrm{Sh}$. At $U_\mathrm{H}$ the unstable cycles disappear via a subcritical Hopf bifurcation and make the time-independent nontrivial states unstable. Further, the unstable nontrivial time-independent states disappear via a pitchfork bifurcation at $U_\ast$, although this does not influence the observable stable regimes of the system. Further increase of $U$ does not lead to any qualitative changes in the system dynamics.}
\label{fig8}
\end{figure*}

\subsection{Dynamics of reduced model (\ref{eq4-09})--(\ref{eq4-10})}
System (\ref{eq4-09})--(\ref{eq4-10}) admits two time-independent solutions: the trivial ground state $A=B=0$ and a nontrivial state
\begin{equation}
A_0=\pm\sqrt{1-\frac{U^2}{K}}\,,\qquad
B_0=-\frac{U}{K}A_0\,.
\label{eq4-11}
\end{equation}
The nontrivial state apparently exists for
\[
U<U_\ast=\sqrt{K}
\]
and disappears via a subcritical pitchfork bifurcation at $U_\ast$.

The stability analysis for the trivial state can be performed analytically (see Appendix for a detailed derivation) and yields the exponential growth rates of linear perturbations
\begin{equation}
\mu_{1,2}=\frac{1-K}{2}\pm\sqrt{\frac{(1+K)^2}{4}-U^2}\,.
\label{eq4-12}
\end{equation}
Exponents $\mu_{1,2}$ are real for
\[
U<U_{\mbox{\small k-f}}=\frac{1+K}{2}
\]
and one of them changes its sign at $U_\ast=\sqrt{K}$ (Fig.~\ref{fig7}). For $U>U_{\mbox{\small k-f}}$, the exponents form a complex conjugate pair with real part $\mathrm{Re}(\mu_{1,2})=(1-K)/2$, which is positive for $K<1$ and negative for $K>1$.

Comprehensive bifurcation analysis can be performed for the dynamic system (\ref{eq4-09})--(\ref{eq4-10}) (Appendix). It reveals the possibility of only two scenarios depending on $K$:
\\
{\bf (I)}\ For $K>1$ (Fig.~\ref{fig7}), the time-independent patterns are the only stable states; the nontrivial solutions disappear via a pitchfork bifurcation at $U=U_\ast$.
\\
{\bf (II)}\ For $K<1$, the sequence of bifurcations is more sophisticated but still universal (see Fig.~\ref{fig8}).
\\
For a Hopf bifurcation the threshold can be calculated analytically:
\[
U_\mathrm{H}=\sqrt{K(K+2)/3}\,.
\]

One can see, that the picture of bifurcation scenarios of the reduced model resembles that of the original model. The only qualitative difference is the lack of a slow decline of the real part of the exponential growth rate of the linear perturbations, which can be seen in Fig.~\ref{fig3}d with a gray line crossing the abscissa at $\mathbf{H_\mathrm{sup}}$. This decline leads to the decay of the oscillatory regime for increasing $u$; in the reduced model, this decay turns out to be not represented, and oscillatory regime in scenario~II persists for large $U$. Otherwise, the scenario~II in the reduced model resembles all fine features of the scenario~II in the original system. Thus, the model reduction firstly validates the completeness of the reported picture of scenarios in the original system and secondly sheds light onto the nature of the bifurcation scenarios in the original model. The underlying difference between two scenarios turns out to be quite simple; the pair of perturbation modes of the trivial state becomes coupled by advection $u$ and converges as $u$ increases (see the upper graphs in Fig.~\ref{fig7}). When the sum of the real parts of the exponential growth rates of modes is negative ($1-K<0$), the convergence leads to a plain decay of both modes. When the sum is positive ($1-K>0$), a reacher sequence of bifurcations must occur near $U_\ast$.

\section{Conclusion}
We have considered the effect of an imposed advection on the localized patterns excited in the modified Kuramoto--Sivashinsky equation (with symmetry $\theta\leftrightarrow-\theta$) under frozen parametric disorder. Firstly, we have revealed that the system evolution is infinitely smooth in time even though the frozen parametric noise is $\delta$-correlated in space. Secondly, while all the stable regimes in the advection-free system are time-independent, stable oscillatory regimes may appear in the presence of advection. Thirdly, two scenarios of advective suppression (``washing-out'' from the excitation centres) of localized patterns have been found with numerical simulation. In scenario~I, the stable patterns remain time-independent for all $u$ and disappear via a pitchfork bifurcation. In scenario~II, a strong advection results in appearance of a stable oscillatory regime, which arises via a tangent bifurcation of cycles. In certain range of advection strength $u$, one observes hysteresis between stable oscillatory and time-independent regimes. For a stronger advection the time-independent regimes become unstable via a subcritical Hopf bifurcation and the oscillatory regime becomes the only attractor in the system. With the further increase of the advection strength, the oscillating regime gradually decays until it disappears via a Hopf bifurcation. Fourthly, we have derived a two-mode reduced model and performed its comprehensive bifurcation analysis, which has yielded that only two reported bifurcation scenarios are possible in the system and revealed the underlying difference between them. With this analysis one can see, that in spite of a seeming complexity of the second scenario, a pair of these scenarios is the simplest possible picture for the effect of the advection, which couples two originally-monotonous modes of perturbations of the trivial state, on the dynamics of a system of the given kind.

\appendix
\section{Bifurcation analysis for the reduced model}
In this appendix section we derive all possible time-independent solutions, their stability properties and local bifurcations for the reduced model (\ref{eq4-09})--(\ref{eq4-10}). Additionally, limitations on the nonlocal objects---limit cycles---are derived on the basis of the Bendixson-Dulac criterion. With this information derived, one can deduce the picture of bifurcation scenarios for the reduced system.

Let us consider the stability properties of the trivial state. The exponential growth rate $\mu$ of linear perturbations is determined by the characteristic equation
\[
\left\|\begin{array}{cc}
  1-\mu & U \\[5pt]
  -U & -K-\mu
\end{array}\right\|=\mu^2-(1-K)\mu+U^2-K=0\,.
\]
This equation yields
\[
\mu_{1,2}=\frac{1-K}{2}\pm\sqrt{\frac{(1+K)^2}{4}-U^2}\,.
\]
Exponents $\mu_{1,2}$ are real for
\[
U<U_{\mbox{\small k-f}}=\frac{1+K}{2}
\]
and one of them changes its sign at $U_\ast=\sqrt{K}$ (Fig.~\ref{fig7}). For $U>U_{\mbox{\small k-f}}$, the exponents form a complex conjugate pair with real part $\mathrm{Re}(\mu_{1,2})=(1-K)/2$, which is positive for $K<1$ and negative for $K>1$.

The perturbations $(A_1,B_1)$ of the nontrivial time-independent state (\ref{eq4-11}) are governed by the equation system
\begin{eqnarray}
&&\dot{A}_1= A_1+UB_1-3A_0^2A_1-3A_0A_1^2-A_1^3,
\label{eq4-13}\\[5pt]
&&\dot{B}_1= -UA_1-KB_1\,.
\label{eq4-14}
\end{eqnarray}
The linear stability of $(A_0,B_0)$ is determined by the characteristic equation
\[
\begin{array}{l}
\left\|\begin{array}{cc}
  1-3A_0^2-\mu & U \\[5pt]
  -U & -K-\mu
\end{array}\right\|=
\\[20pt]
\qquad\displaystyle
\mu^2-\left(\frac{3U^2}{K}-2-K\right)\mu +2\left(K-U^2\right)=0\,,
\end{array}
\]
which yields
\begin{eqnarray}
&&\hspace{-10pt}
\mu_{3,4}=-\left(1+\frac{K}{2}-\frac{3U^2}{2K}\right)
\nonumber\\[5pt]
&&\quad{}
\pm\sqrt{\left(1+\frac{K}{2}-\frac{3U^2}{2K}\right)^2-2\left(K-U^2\right)}\,.
\label{eq4-15}
\end{eqnarray}

The nontrivial state is always stable at $U=0$. As one can see from Eq.~(\ref{eq4-15}), $\mu_{3,4}$ cannot change its sign where it is real-valued, since the state exists for $U^2<K$. Hence, the condition of the stability change is $\mathrm{Re}(\mu_{3,4})=-(1+K/2-3U^2/(2K))=0$ and it occurs via a Hopf bifurcation. The condition yields
\[
U_\mathrm{H}=\sqrt{\frac{K(K+2)}{3}}\,,
\]
which is smaller that $U_\ast$ for $K<1$ and bigger otherwise. Thus, the bifurcation of stability change occurs only for $K<1$; for $K>1$ the time-independent nontrivial state is stable until in disappears at $U_\ast$ (Fig.~\ref{fig7}).

For a Hopf bifurcation at $U_\mathrm{H}$, from Eq.~(\ref{eq4-15}), one can calculate frequency \[
\omega=\mathrm{Im}(\mu_{3})=\sqrt{\frac{2}{3}K(1-K)}\,,
\]
and the time-independent solution amplitude $A_0=\sqrt{(1-K)/3}$. For this bifurcation one can employ the standard multiple scale method and after laborious but straightforward calculations find the amplitude equation for perturbations $(A_1,B_1)=C(t)\,(K+i\omega,-U_\mathrm{H})e^{i\omega t}+c.c.$, where $c.c.$ stands for complex conjugate;
\begin{equation}
\dot{C}=\frac{\mathrm{d}\mu_3}{\mathrm{d}U}(U-U_\mathrm{H})\,C+\alpha C\left|C\right|^2\,,
\label{eq4-16}
\end{equation}
\[
\alpha=-\frac{6(K^2+\omega^2)(K+i\omega)^2(2K-i\omega)}{K\big(K^2+\omega^2-(K+i\omega)^2\big)}\,.
\]
Employing the expression for $\omega(K)$, one can calculate coefficient $\alpha$ and see that it possesses a positive real part for $0<K\le 1$, where the bifurcation can occur. Thus, this Hopf bifurcation is always subcritical; the limit cycle is unstable and exists for $U<U_\mathrm{H}$, at the bifurcation point the unstable limit cycle collapses onto the time-independent state and makes it unstable (Fig.~\ref{fig7}).

At $U=0$, the dynamics of $A$ and $B$ is decoupled and no cycle is possible (see Eqs.~(\ref{eq4-09})--(\ref{eq4-10})). Hence, the unstable cycle collapsing onto the time-independent state must appear somewhere between $U=0$ and $U_\mathrm{H}$. Let us consider the possibility of the appearance of cycles in the system under consideration. To employ the Bendixson-–Dulac criterion, one has to calculate the divergence of the phase flux $\vec{F}$;
\[
\nabla\cdot\vec{F}=\frac{\partial\dot{A}}{\partial{A}}+\frac{\partial\dot{B}}{\partial{B}}
=1-K-3A^2\,.
\]
According to the Bendixson-–Dulac criterion, the integral of the divergence over the phase plane domain bounded by a cycle must be zero, {\it i.e.}, both the domains of positive and negative values of the divergence must be present within the cycle. For $K>1$, the divergence is negative everywhere and no cycle can appear in the system. For $K<1$, the divergence changes its sign on
\[
A_\mathrm{BD}=\pm\sqrt{\frac{1-K}{3}}\,,
\]
and any cycle must cross at least one of these two lines.

Let us discuss the case of $K<1$ in detail. For $U>U_\ast$ the only fixed point in the system is the unstable trivial solutions, simultaneously the nonlinear term $-A^3$ makes the trajectories to go from infinity towards some finite vicinity of the phase plane origin; therefore, there must be at least one attracting limit cycle in the system and, as shown above, it must cross $A=A_\mathrm{BD}$. Since there is no limit cycle for $U=0$, this cycle must appear somewhere for $0<U<U_\ast$. The simplest possible option is a tangent bifurcation for a pair of stable and unstable cycles; the bifurcation point $U_\mathrm{T}$ (see Fig.~\ref{fig8}) must lie between $0$ and $U_\mathrm{H}$, where the unstable limit cycle disappears. Recall, at $U_\mathrm{H}$ a symmetric pair of unstable limit cycles collapses on the fixed points; the only way, how a large unstable cycle appearing at $U_\mathrm{T}$ can be split into two smaller cycles is the formation of homoclinic loops at certain point $U_\mathrm{Sh}$ (see Fig.~\ref{fig8}). This bifurcation of the formation of homoclinic loops differs from the standard case of the Shil'nikov's theorem~\cite{Shilnikov-1966} due to the system symmetry $(A,B)\leftrightarrow(-A,-B)$. The described bifurcation sequence presented in Figs.~\ref{fig7}, \ref{fig8} is the simples scenario admitted by the analytically tractable features we derived in this section for system (\ref{eq4-09})--(\ref{eq4-10}). The numerical simulation of the reduced system for the entire parameter space reveals that this simplest scenario is the only one occurring in the system for $K<1$.

According to the derivations of this section, the reduced system can follow strictly two scenarios: Scenario~I for $K>1$ and Scenario~II for $0<K<1$.

\ack{
The author is thankful to A.\ V.\ Dolmatova and L.\ S.\ Klimenko for useful comments and discussions.
The work has been supported
 by the Russian Science Foundation (Grant no.\ 14-21-00090).}

\end{document}